\documentclass[epj]{svjour}
\usepackage{graphics}
\usepackage{epsfig}
\newcommand{\ds}{\displaystyle}
\newcommand{\dsf}{\ds\frac}
\newcommand{\Tr}{\mbox{Tr}}
\newcommand{\re}[1]{(\ref{#1})}
\newcommand{\no}{\nonumber}
\begin{document}
\title{Electromagnetic form factors of bound nucleons revisited}
\author{Ulughbek T. Yakhshiev \inst{1,}~\inst{2} 
\and Ulf-G. Mei{\ss}ner  \inst{1} \thanks{Address after January 1$^{\rm
    st}, 2003$: Helmholtz Institut f\"ur Strahlen- und Kernphysik (Theorie),
    Universit\"at Bonn, Nu{\ss}allee 14-16, D-53115 Bonn, Germany.}
\and Andreas Wirzba \inst{1}
}                     
%
%
\institute{Institut f\"ur Kernphysik (Theorie), Forschungszentrum J\"ulich, 
D-52425 J\"ulich, Germany 
\and Theoretical Physics Department and Institute of Applied Physics,
National University of Uzbekistan, Tashkent-174,\\ Uzbekistan}
\date{Received: date / Revised version: date}
%
\abstract{
We investigate the possible modifications of the nucleons' electromagnetic
form factors in the framework of a modified Skyrme model allowing for 
nucleon deformation and using realistic
nuclear mass distributions. We show that such effects are small in light nuclei.
\PACS{
      {13.40.Gp}{} \and
      {12.39.Fe}{} \and
      {21.65.+f}{} 
     } 
} 
%


\maketitle

\section{Introduction}
\label{intro}

It is firmly believed that hadron properties must undergo changes in a
baryon--rich or hot environment. Many data from quasi--elastic or deep
inelastic electron scattering off nucleons or nuclei, deeply bound pionic
atoms and from heavy--ion reactions point towards such effects, but despite a
tremendous amount of effort no clear theoretical picture concerning such scale
changes has emerged\footnote{For example, the important question of
  disentangling genuine scale changes from ``standard'' many--body effects
  remains to be solved in a concise manner.}. It is therefore mandatory to
study realistic models of the nucleon (or other hadrons) and to investigate how
certain properties will change at finite baryon density. A popular class of
such models are based on the Skyrme Lagrangian and variations thereof, for
essentially three reasons: First, such models have proven to lead to a fairly
good description of a wide variety of nucleon properties in free space and
secondly, being based on non--linearly interacting pion fields, they are built
from the degrees of freedom directly related to the spontaneous chiral
symmetry breaking that QCD is supposed to undergo.  Third, in contrast to
(most) quark models, the pion cloud contribution of the nucleon is naturally
taken into account.  In earlier studies rather large medium modifications were
found, based on the assumption of a constant background density (homogeneous
nuclear matter), see e.g. Refs.~\cite{Celenza,UGM,Ruiz,LuGEn,lu,musa,prc001}.
In Ref.~\cite{npa002} we had formulated a more realistic version of
such an approach, taking into account not only realistic distributions
of baryon density within light, medium and heavy nuclei, but also
allowing the nucleons to deform under the influence of the ensuing 
baryon density gradients. In particular, it was shown that the popular
concept of a uniform size modification (nucleon swelling or shrinking) 
cannot be maintained in such a picture, i.e. that the influence of the nuclear
medium and the response of the nucleon to it is very probe dependent.
In particular, the scale changes of the isoscalar and isovector
electromagnetic charge distributions depend on the direction
considered (there is an axial symmetry around the direction from the
center of the nucleus to the center of the nucleon at some distance
$R$) giving rise to a small {\em intrinsic}  quadrupole moment. While the
proton's shape changes from oblate to a prolate shape as it is moved
toward the surface of the nucleus, the behaviour of the neutron is
just the opposite. It is therefore of interest to reconsider the 
possible modifications of the nucleons' electromagnetic form factors,
triggered also by the experimental fact that in the absence of neutron
targets one has to use light nuclei like deuterium or $^3$He to
determine the neutron form factors. This will be the topic of the
present paper, where we study these form factors inside $^4$He, which
is the lightest nucleus that can be approximated by a continuous matter
distribution (for a similar calculation including nuclear shell effects,
see \cite{lu}).  Finally, we note that a more systematic approach to tackle
these questions based on effective field theory is not yet available for
nucleons and only in its infancy for pions; for an attempt see e.g.
\cite{MOW} (and references therein).

\medskip\noindent
Our work is organized as follows. In Section~\ref{sec:model} we
briefly review the underlying modified Skyrme model and give formulae
for the electromagnetic form factors for the case of finite baryon
density. In Section~\ref{sec:results}, we show and discuss the
results for the specific case of $^4$He. We refrain from discussing 
other nuclei as done e.g. in Refs.\cite{lu,npa002} since these results 
are genuine and can be scaled easily. We conclude with a short summary 
in Section~\ref{sec:summ}. 

\vfill

\section{Brief description of the model}
\label{sec:model}

\subsection{Lagrangian}
Our starting point is a modified Skyrme Lagrangian in the nuclear
medium~\cite{prc001}
\begin{equation}
\begin{array}{l}
{\cal L}=
\dsf{F_{\pi}^2}{16}\mbox{Tr}\left(\dsf{\partial{U}}{\partial t}\right)
\left(\dsf{\partial{U}^\dagger}{\partial t}\right)-
\dsf{F_{\pi}^2}{16}\alpha_p(\vec r)\mbox{Tr}
(\vec\nabla{U})(\vec\nabla{U^\dagger})\\
\quad \\
\qquad \qquad
+\dsf{1}{32e^2}\mbox{Tr}[L_{\mu},L^{\nu}]^2
+\dsf{F_{\pi}^2 m_{\pi}^2}{16}\alpha_s(\vec r)\mbox{Tr}[U+U^{\dagger}-2]
\label{lag}
\end{array}
\end{equation}
with $F_\pi$ the weak pion decay constant, $m_\pi$ the
pion mass,  $U(\vec r \,)$ parametrizes the
pion fields and $L_\mu = U^\dagger \partial_\mu U$ is the left-handed
current. The Skyrme parameter $e$ (together with the pion
decay constant) will be determined from fitting the
masses of the nucleon and the $\Delta (1232)$. Its value can be 
understood in terms of rho meson exchange.
We use  a deformed ansatz for the pion fields when the Skyrmion
is  located at some distance $\vec R$ from the center of the nucleus\footnote{
For the details on the geometry and the justification of this ansatz, see
\cite{npa002}.}
\begin{eqnarray}
U({\vec r})&=&\exp\left[i\vec\tau \cdot \vec N(\Theta(\theta),\varphi)\,
F(r,\theta)\right]~,\\
\vec N&=&\{\sin\Theta(\theta)\cos\varphi,\sin\Theta(\theta)\sin\varphi,
\cos\Theta(\theta)\}~,\\
F(r,\theta)&=& 2\arctan\biggl\{\left(\dsf{r_S^2}{r^2}\right)
[1+\gamma_1\cos\theta  \label{ansatzF}\nonumber\\
&+& \gamma_2\cos^2\theta+\gamma_3\cos^3\theta+\dots]\biggr\}~,\\
\Theta(\theta)&=&\theta+\delta_1\sin 2\theta+\delta_2\sin 4\theta
 +\delta_3\sin6\theta+ \ldots~.
\label{ansatz}
\end{eqnarray}
Here, $F(r,\theta)$ is the profile function of the deformed
Skyrmion and  $r_S$, $\gamma_1$, $\gamma_2$, $\gamma_3$, $\ldots$ and
$\delta_1$, $\delta_2$, $\delta_3$, $\ldots$ are variational
(or, respectively, deformation) parameters which are determined
from minimizing the Skyrmion energy for a given background baryon
density. The density dependence is contained in the medium
functionals $\alpha_s (\vec r)$ and $\alpha_p (\vec r)$, 
\begin{eqnarray}
\alpha_p(\vec r )&=& 1-\frac{4\pi c_0 \rho(\vec r) /\eta}
{1 +g_0' 4\pi c_0 \rho(\vec r)/\eta}\,,\\
\alpha_s(\vec r )&=&1 - 4\pi \eta b_0 \rho(\vec r)/m_\pi^2\,.
\end{eqnarray}
Here $\eta=1 +m_\pi/m_N \sim 1.14$ is a kinematical factor, $m_N = 938\,{\rm
  MeV}$ the mass
of the nucleon, $g_0'=1/3$ the Migdal parameter which takes into account the
short-range correlations, and $b_0=-0.024\,m_\pi^{-1}$ and
$c_0=0.21\,m_\pi^{-3}$ are 
empirical parameters which can be taken from the analyses
of low-energy pion-nucleus scattering data~\cite{ericson}. For more details see
Ref.~\cite{npa002}. Since Lorentz invariance is broken
at finite density, the time and space derivatives acting on the
pion fields have different prefactors.  To obtain states with
definite spin and isospin, one has to perform adiabatic rotations
and quantization of these. In  Ref.~\cite{npa002}, we have calculated
the nucleon mass and other static properties for nucleons inside light,
medium and heavy nuclei, based on realistic nuclear density distributions within
the nuclei considered. For example, the decrease of the nucleon mass
came out considerably smaller than in earlier studies where uniform
baryon matter densities where assumed. It was also shown that the
concept of a uniform swelling of nuclear sizes in the medium is too simple
to be a realistic picture, in fact the modifications for the baryon matter
distribution within a nucleon or the scale changes of the various
electromagnetic radii all turned out to be different. In this paper,
we extend these considerations to the nucleons form factors at small
and intermediate momentum transfer.

\subsection{Electromagnetic form factors}
The electric and magnetic form factors of the nucleon are defined 
through the expressions
\begin{eqnarray}
G_E (q^2)&=&\ds\int d^3r \, e^{i\vec q\cdot\vec r}j^0(r)~,\nonumber\\
G_M (q^2)&=&m_N\ds\int d^3r \, e^{i\vec q\cdot\vec r}[\vec r\times \vec j(r)]~,
\label{ffdef}
\end{eqnarray}
where $q^2$ is the momentum transfer squared, $j^0$ and $\vec j$ correspond
to the time and the space components of the properly normalized sum
of the baryonic current $B_\mu$ 
and the third component of the isovector current $\vec V_\mu$,
i.e.
\begin{eqnarray}
B_\mu &=& \dsf{1}{24\pi^2}\, \varepsilon_{\mu\nu\alpha\beta}
\, \Tr \, L^\nu L^\alpha L^\beta\,,\\
V_{\mu}^{(3)} &=&-\dsf{i F_{\pi}^2}{16} C_{\mu}\Tr\,
\tau_3(L_{\mu}+R_{\mu}) \nonumber \\  &+&
\!\dsf{i}{16{e}^{2}}\Tr\,\tau_3\Bigl\{\bigl[L_{\nu},[L_{\mu},L_{\nu}]\bigr]\!+\!
\bigl[R_{\nu},[R_{\mu},R_{\nu}]\bigr]\Bigr\}\,\!,\\
R_{\mu}&=&U\partial_{\mu}U^+;\quad
C_{\mu}=\left\{\begin{array}{ccl}
1 &, &\mbox{$\mu=0$}\,,\\
\alpha_p&, & \mbox{$\mu=1,2,3$}\,,\\
\end{array}\right.
\label{currents}
\end{eqnarray}
with $\varepsilon_{\mu\nu\alpha\beta}$ the totally antisymmetric
tensor in four dimensions and Tr denotes the trace in SU(2) flavor space.
Evaluating these current operators between appropriate nucleon states
as described in \cite{npa002}, one obtains
the  electromagnetic form factors of the nucleon.
For the problem at hand, it is advantageous to
expand the plane wave factor in the expressions Eqs.\re{ffdef},
\begin{equation}
e^{i\vec q\cdot\vec r}=4\pi\ds\sum_{l=0}^{\infty}\sum_{m=-l}^l
i^lY_l^{m}(\theta_q,\varphi_q)Y_l^{m*}(\theta_r,\varphi_r)j_l(qr)~,
\end{equation}
in terms of spherical harmonics $Y_l^m$ and spherical
Bessel  functions 
$j_l$. In this way
we get the final expression for the electromagnetic form factors
\begin{equation}
G_a^b(q^2)=\ds\sum_l i^l\sqrt{2l+1}P_l(\cos\theta_q){G}_a^{b,l}(q^2)~,
\label{ffull}
\end{equation}
where the label $a$ stands either 
for electric $(E)$ or magnetic $(M)$ form factors,
$b$ stands either for isoscalar $(S)$ or isovector $(V)$ form factors, and
the $P_l$ are Legendre polynomials. After angular integration in 
momentum space\footnote{We note that in the present case the
incoming beam direction is in general not the  $z$--direction of the
coordinate system, since the latter is fixed by the direction
of ${\bf R}$.}, the moduli of the form factors satisfy the simple rule
\begin{equation}
\ds\int d\Omega_q~|G_a^b (q^2)|^2 = 4\pi\sum_{l=0}^\infty (G_a^{b,l}(q^2))^2~. 
\end{equation}
The corresponding partial form factors have the form
\begin{eqnarray}
G_E^{S,l}(q^2)&=&-\dsf{\sqrt{2l+1}}{\pi}
\int\limits_0^{\infty}r^2dr
\int\limits_0^\pi\sin\theta d\theta \nonumber\\
&\times&\left\{j_l(qr) P_l(\cos\theta)F_r\Theta_\theta
\dsf{\sin^2F}{r^2}\dsf{\sin\Theta}{\sin\theta}\right\}~,
\nonumber \\
G_E^{V,l}(q^2)&=&~\dsf{\sqrt{2l+1}~\pi}{4 I_{\Omega\omega}^{(33)}}
\int\limits_0^{\infty}r^2dr\int\limits_0^\pi\sin\theta
d\theta \nonumber\\
&\times&
\biggl\{j_l(qr) P_l(\cos\theta)
\sin^2\Theta\dsf{\sin^2F}{r^2}\Bigl[F_\pi^2 r^2 \nonumber\\
&\quad& +   \frac{4}{e^2}\left(r^2F_r^2+F_\theta^2+
\Theta_\theta^2\sin^2F\right)\Bigr]\biggr\}~,
\nonumber\\
G_M^{S,l}(q^2)&=&-\dsf{\sqrt{2l+1}~m_N}{4\pi I_{\Omega\omega}^{(33)}}
\int\limits_0^{\infty}r^2dr\int\limits_0^\pi\sin\theta
d\theta \nonumber\\ &\times&
\biggl\{j_l(qr) P_l(\cos\theta)
F_r\Theta_\theta\sin^2{F}\sin\Theta\sin\theta\biggr\}~,
\nonumber\\
G_M^{V,l}(q^2)&=&~\frac{1}{3}{\sqrt{2l+1}~\pi m_N}
\int\limits_0^{\infty}r^2dr\int\limits_0^\pi\sin\theta
d\theta \nonumber\\
&\times&
\biggl\{j_l(qr) P_l(\cos\theta)
\sin^2\Theta\dsf{\sin^2F}{r^2}\Bigl[F_\pi^2 r^2\alpha_p \nonumber\\&\quad&
+\frac{4}{e^2}\left(r^2F_r^2+F_\theta^2+
\Theta_\theta^2\sin^2F\right)\Bigr] \biggr\}~,
\label{FFs}
\end{eqnarray}
where $F_r=\partial F/\partial r$, $F_\theta= \partial F/\partial \theta$,
$\Theta_\theta= \partial \Theta/\partial\theta$ are partial derivatives and
\begin{eqnarray}
I_{\Omega\omega}^{(33)}&=&\dsf{\pi}{4}
\int\limits_0^{\infty}dr\int\limits_0^\pi\sin\theta
d\theta
\biggl\{\sin^2{\Theta}\sin^2{F} \nonumber\\
&\times & \, \Bigl[F_\pi^2 r^2+\frac{4}{e^2}\left(r^2F_r^2+F_\theta^2+
\Theta_\theta^2\sin^2F\right)\Bigr]\biggr\}~,
\end{eqnarray}
is a moment of inertia. The details of the quantization procedure
to obtain states with good spin and isospin quantum numbers from
the deformed topological soliton are spelled out  in 
\cite{npa002}.

\medskip\noindent
The nucleon form factors are defined as
\begin{equation}
G_{E,M}^{({p\atop{n}})} (q^2) =\dsf{1}{2}\, 
\left(G_{E,M}^S (q^2) \pm G_{E,M}^V  (q^2) \right)~.
\end{equation}
We note that the isovector magnetic form factor explicitly depends on the
medium functional $\alpha_p(r,\theta,R)$. We now want to consider
the possible modifications of these form factors within $^4$He.
As in the previous work~\cite{npa002} the density is parametrized 
as\footnote{Note that we correct for an typographical error that
appeared in Eq.~(18) of Ref.\cite{npa002}.}
\begin{equation}
\rho(r)=\left( \dsf{3}{4}\right)\,
\dsf{2}{\pi^{3/2}r_0^3}
\left[1+\dsf{A-2}{3}
\left(\dsf{r^2}{r_0^2}\right)\right]
\exp
\left\{-\dsf{r^2}{r_0^2}\right\}~,
\label{density}
\end{equation}
where the prefactor $3/4$ accounts
for the fact that we single out one nucleon from the background of the
others. The parameter $r_0=1.31$~fm corresponds to $^4$He~\cite{akhiezer}. 
We are now in the position to study the nucleons' electromagnetic
form factors within this light nucleus. 

\section{Results and discussion}
\label{sec:results}

\begin{figure*}[htb]
 \vspace{4.5cm}
\includegraphics{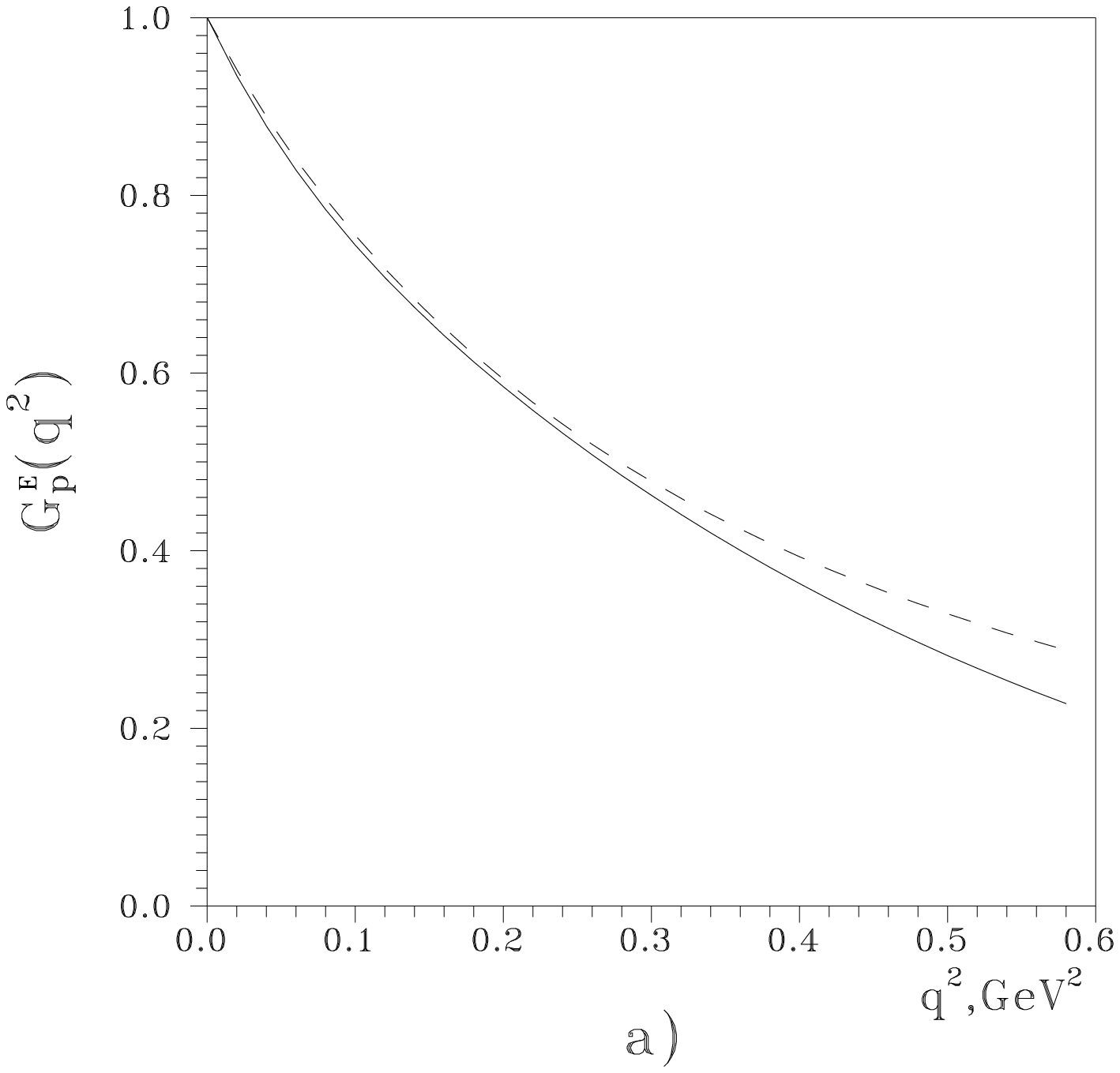}
\includegraphics{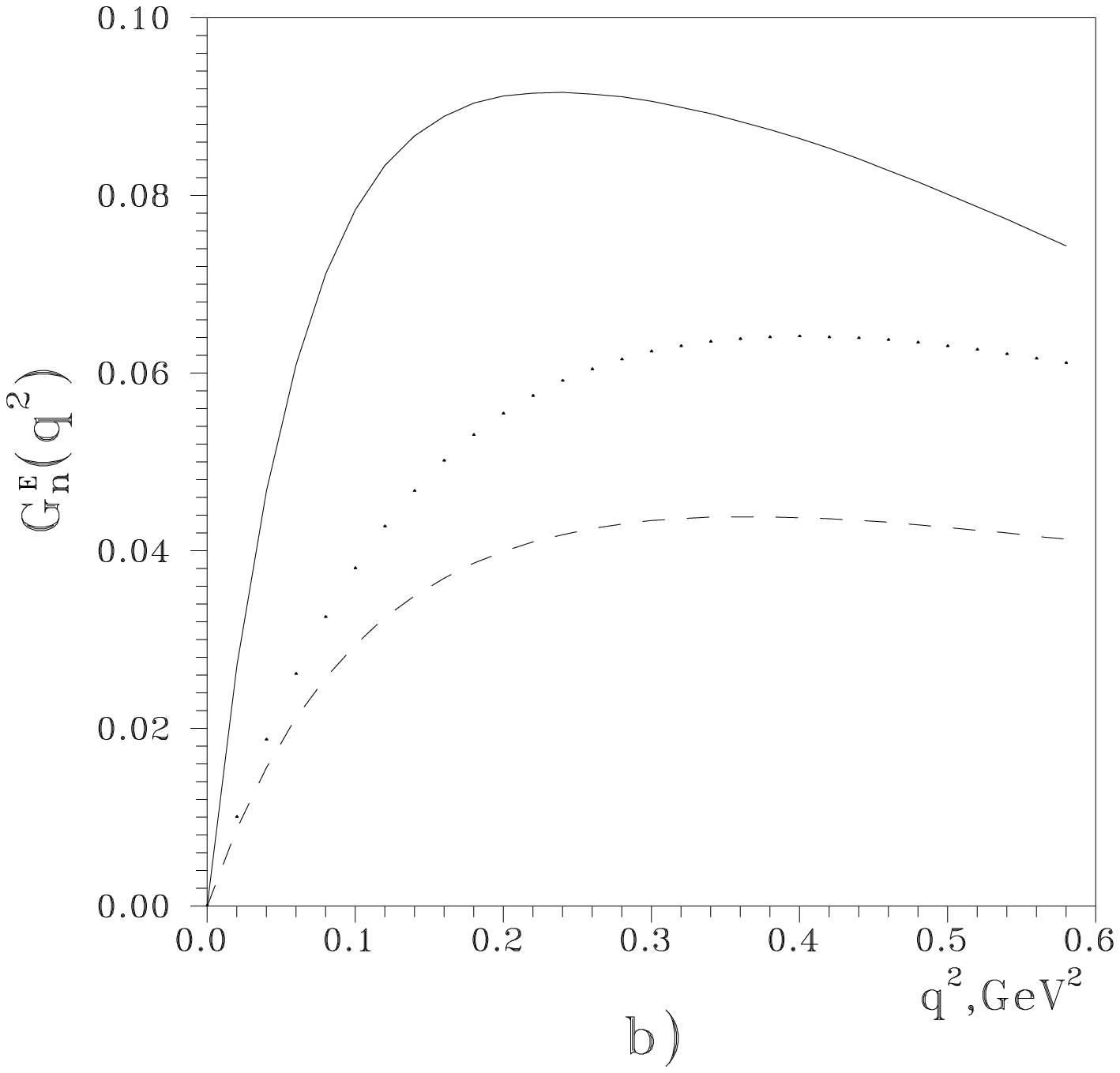}
\vspace{15mm}
 \caption{\it
      Electric proton (a)  and electric neutron (b) 
      form factor. Solid (dashed) lines: our model (dispersion-theoretical
      analysis of \protect \cite{MMD}). The dotted line in $G_E^n (q^2)$
      represents the parametrization Eq.(\ref{GalsterM}).
    \label{fffree} }
\end{figure*}

Our input parameters are the same as in the Ref.~\cite{npa002},
i.e. we use $F_\pi$=108~MeV and $e$=5.265 (to fit the nucleon and the delta
mass). Before discussing any possible medium modifications, we have to
consider the electromagnetic form factors in free space. In Fig.~\ref{fffree}
we show the charge form factors of the proton and the neutron in comparison
to the dispersion-theoretical results of Ref.~\cite{MMD} and also to the
recent Galster--like parameterization of $G_E^n(q^2)$ \cite{herberg}
\begin{equation}\label{GalsterM}
G_E^n(q^2) = -\frac{\mu_n \, \tau}{1 + 3.4\, \tau}\,  
\left( 1 + {q^2 \over 0.71\,{\rm GeV}^2} \right)^{-2}~,
\end{equation} 
with $\tau = q^2 / 4m_N^2$. The model predictions are
in fair agreement with the data, more precisely with the 
phenomenological fits. A similar statement holds for the 
momentum dependence of the proton and neutron magnetic form factors, 
not shown here. Note that we display 
the form factors only for momentum transfer
squared $q^2 \le 0.6$~GeV$^2$ for two reasons. First, the model does not
contain vector mesons which start to be relevant at a typical scale
$m_\rho^2 = (0.77~{\rm GeV})^2 \simeq 0.6~{\rm GeV}^2$ (for a review see
\cite{ulfrev}) and second, boost effects can not be completely ignored
any more for these momentum transfers
(see e.g. \cite{holzwarth} for a discussion on this point).
The magnetic moments come out too small, as it is well--known
in such type of models. We have $\mu_p = 1.93$ and $\mu_n =-1.20$ in units
of nuclear magnetons. Note, however, that the ratio $|\mu_p / \mu_n| = 1.61$ 
is close to the empirical value of $1.46$. 
Other static properties are given in \cite{npa002}.

\medskip\noindent
We now consider the medium modifications due to the finite baryon density
within the $^4$He nucleus. In Table~\ref{tab:prop} we give the modifications
of the nucleon mass, of the proton and neutron magnetic moments and of the
variational parameter $r_S$ characterizing the Skyrmion size~\footnote{This
  variational parameter should not be mixed up with the (isoscalar) 
  r.m.s.\ radius of the
  nucleon, see Ref.\cite{npa002}.}, cf. Eq.~(\ref{ansatzF}), for various
distances from the center of the nucleus. We note that the magnetic moments
are changed by less than 5.5\%. In particular, while we find a mild
suppression in the center of the nucleus, at distances above 1~fm the moments
are in fact slightly enhanced in magnitude, very different from the behavior
of the nucleon mass.  The changes in $r_S$ are of a similar size. Note,
however, that the magnitude of $r_S$ decreases monotonically from the center
of the nucleus to its surface.  These results further demonstrate that the
notion of a uniform swelling (or shrinking) of the various nucleon sizes or
properties is ruled out by such type of realistic model.
\renewcommand{\arraystretch}{1.2}
\begin{table}[htb]
\caption{Properties of nucleons in $^4$He and in free space
(last row). $R$ is the  distance
between the centers of the nucleon and the $^4$He nucleus;
$r_S$ is a variational parameter as explained in the text.
\label{tab:prop}}
\begin{center}
\begin{tabular}{|c|cccc|}
\hline
$R$ [fm ]& $r_S$ [fm] & $\mu_p$ [n.m.] & $\mu_n$ [n.m.]& $m_N/m_N^{\rm free}$\\
\hline
  0  & 0.627 & 1.883 & -1.137 & 0.817 \\ 
0.25 & 0.626 & 1.885 & -1.140 & 0.823 \\
0.50 & 0.624 & 1.892 & -1.150 & 0.837 \\
0.75 & 0.621 & 1.907 & -1.168 & 0.860 \\
1.00 & 0.619 & 1.930 & -1.195 & 0.887 \\
1.25 & 0.618 & 1.956 & -1.224 & 0.915 \\
1.50 & 0.617 & 1.975 & -1.245 & 0.941 \\
1.75 & 0.615 & 1.982 & -1.251 & 0.961 \\
2.00 & 0.611 & 1.975 & -1.243 & 0.978 \\
2.25 & 0.607 & 1.962 & -1.230 & 0.988 \\
2.50 & 0.604 & 1.950 & -1.217 & 0.994 \\
\hline 
-& 0.600 & 1.932 & -1.197 &  1  \\   
\hline
\end{tabular}
\end{center}
\end{table}

\begin{figure*}[htb]
 \vspace{9.0cm}
\includegraphics{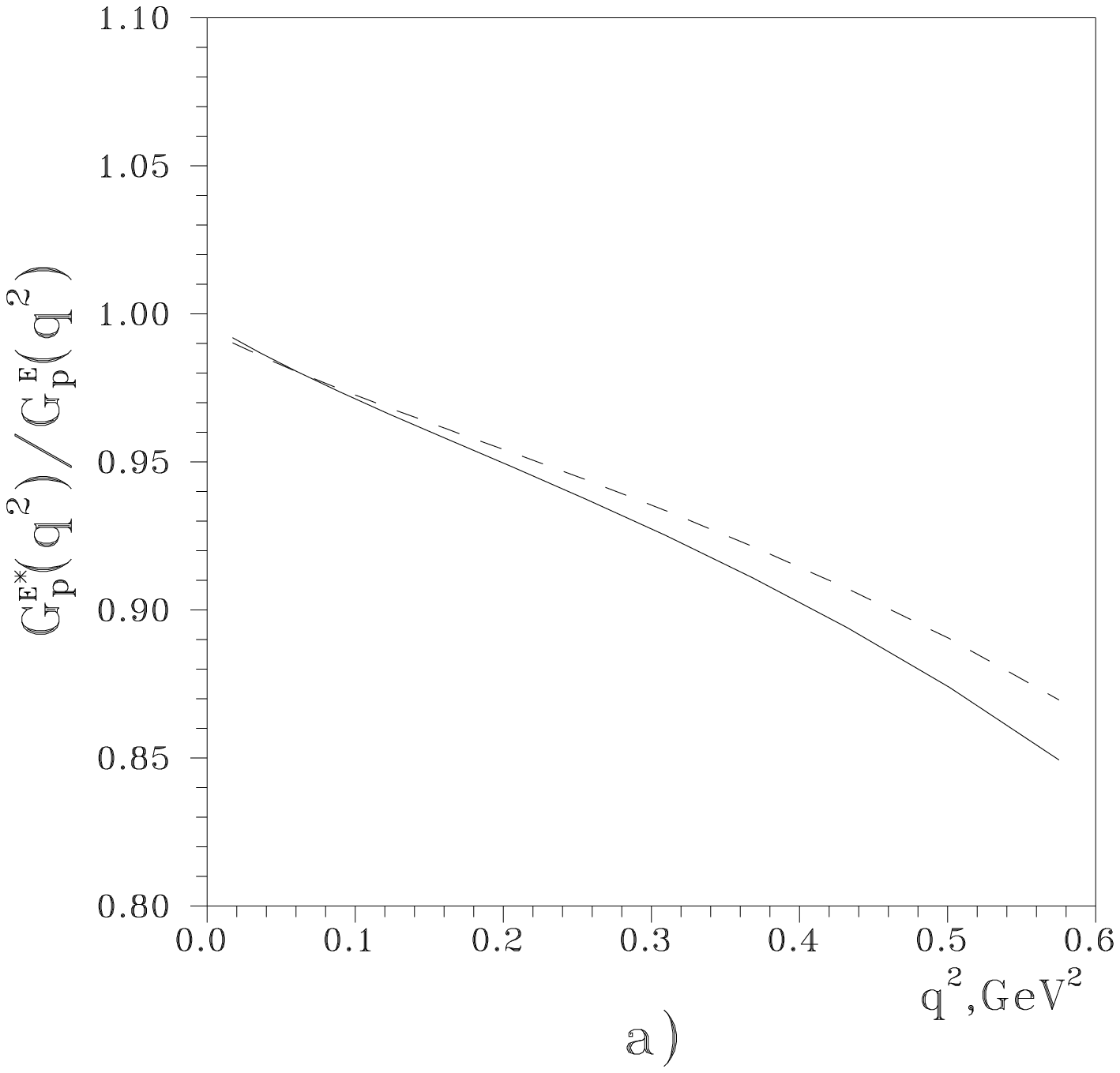}
\includegraphics{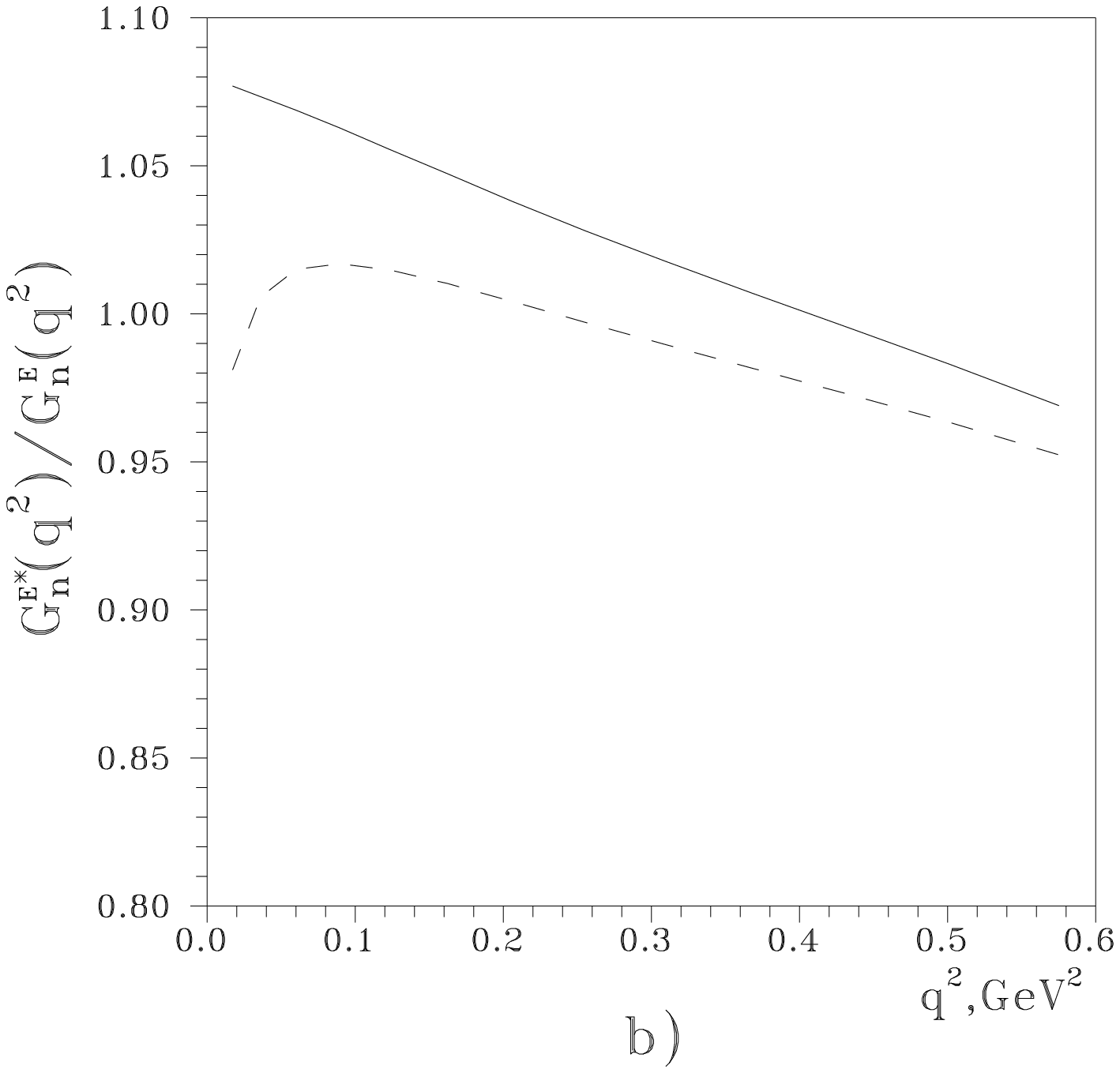}
\includegraphics{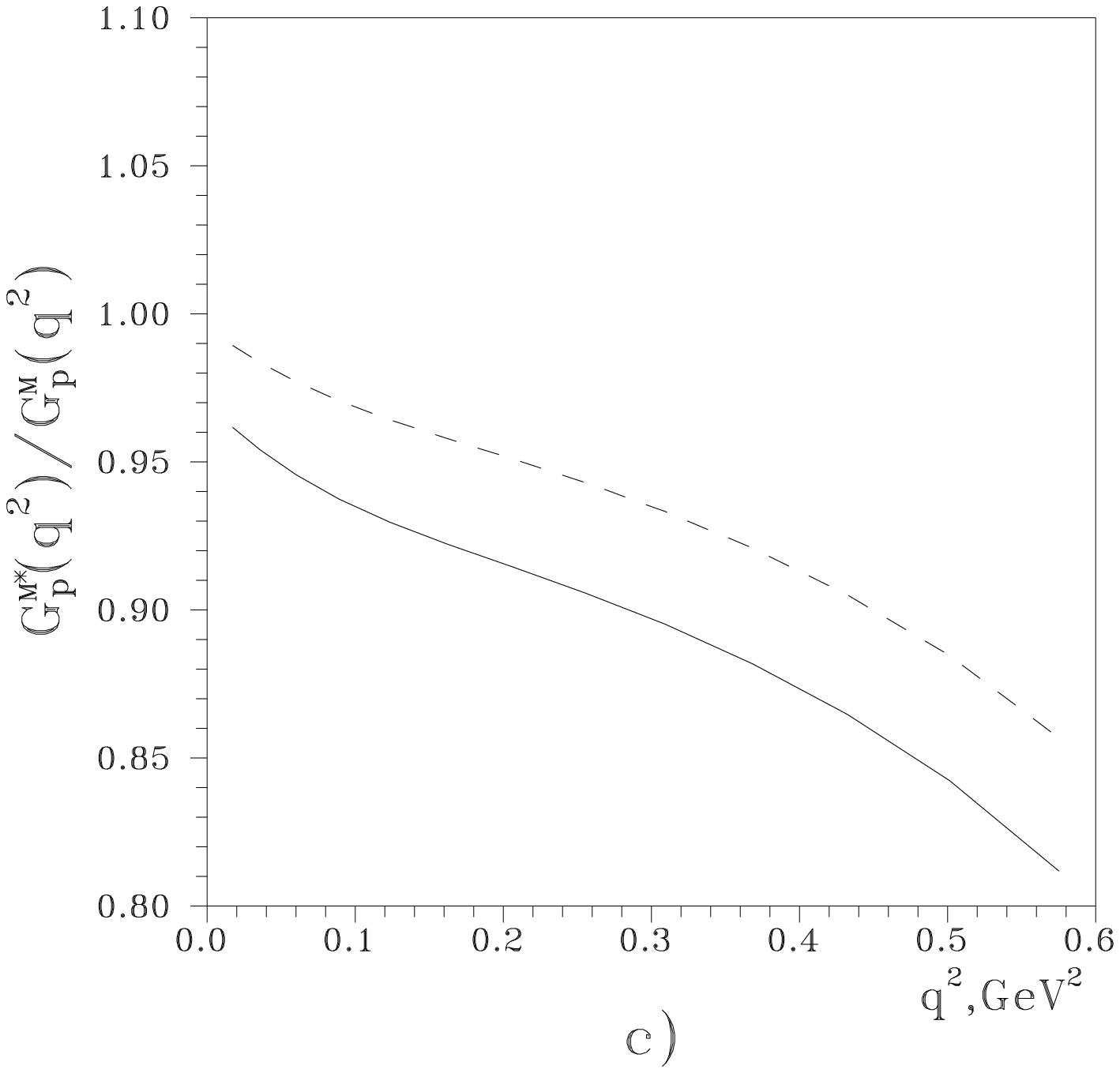}
\includegraphics{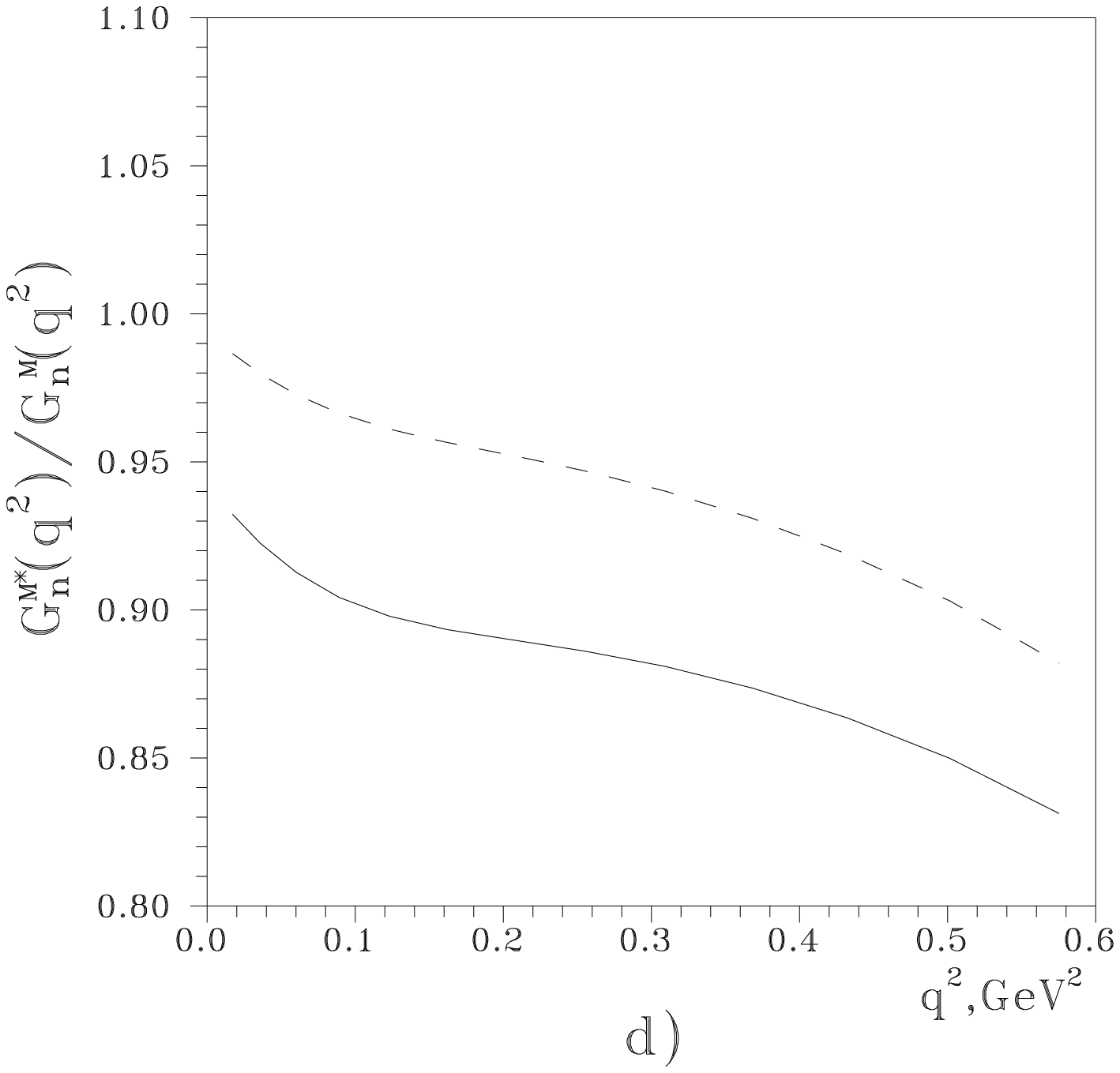}
\vspace{15mm}
 \caption{\it
      Form factors normalized to their free space values for two densities within
      the $^4$He nucleus. The distance of the center of the nucleon from the center
      of the nucleus is 0~fm (solid lines) and 1~fm (dashed lines).
      Panels a), b) c) and d) give the 
      proton electric, neutron electric, proton magnetic and neutron magnetic
      form factor, in order. 
    \label{ffbound} }
\end{figure*}

\noindent
In Fig.~\ref{ffbound} we show the $l=0$ projections of the
four nucleon form factors in the medium normalized to their 
free space values for two different
densities, that is various distances from the center of the
nucleus. These are $R = 0\,(1)\,$ fm corresponding to a residual 
density, cf. Eq.~(\ref{density}),
of $0.7 \, (0.55)\,\rho_0$, with $\rho_0 = 0.17\,$fm$^{-3}$ the nuclear
matter density. 
We note that the medium modifications are  small
for $q^2 \leq 0.6\,$GeV$^2$, i.e. for momentum transfers where the
model can be considered realistic. These changes stay below 20\%
for all form factors. They are in particular small for $G_E^n (q^2)$,
which is often considered to be the most sensitive quantity with respect to
such medium modifications. It is interesting to consider the isospin 
basis. While for the electric case, the isovector piece shows a stronger
medium dependence than the isoscalar one, the magnetic isovector and
isoscalar form factors exhibit approximately the same suppression
for the range of momentum transfers considered here. This latter trend
was also found in earlier calculations \cite{UGM}.
The partial form factors for $l \ge 1$, which we do not show,
are very small. The results for the proton charge and magnetic form factor are
comparable to the ones obtained in
\cite{lu} in the framework of a quark-meson model and employing shell 
like nuclear density distributions, but differ in finer details like
the magnitude of the modifications. Note, however, that these authors
apply their model to a considerably larger range of momentum transfer.
Clearly, our results are also consistent
with the limits obtained from electron scattering data based on the
$y$--scaling hypothesis \cite{sick}.

\section{Summary and outlook}
\label{sec:summ}

In this paper, we have considered the possible medium modifications of
the nucleons electromagnetic form factors in light nuclei,
specifically within $^4$He. We have used an extended Skyrme model
allowing for deformations of the nucleons immersed in the nuclear
medium and applying realistic nuclear matter distributions. We find
small medium renormalizations, quite different from the ones obtained
using a homogeneous nuclear matter background. 
In particular,
these form factors are not uniformly changed.
This indicates that the concept of a uniform swelling or shrinking of the
nucleon sizes cannot be maintained. These results are also
consistent with the ones found in similar type of models when proper care
is taken about the distribution of the nuclear density distributions.
Beside the small effects considered here, there are also many--body 
effects that lead to in--medium nucleon changes. These are expected
to be significant only in heavier nuclei. From the results presented
here, we must conclude that the extraction of the neutron charge
form factor from light nuclei at low and intermediate momentum
transfer is not sensitive to such effects, given the presently
achieved experimental accuracy.

\section*{Acknowledgements}

We gratefully acknowledge helpful discussions with 
Yousuf Musakhanov and Abdulla Rakhimov.
We thank Hartmut Schmieden for a useful communication.
The work of U.Y. has been supported by an INTAS YS fellowship N$^\circ$00-51 and
SCOPES grant N$^\circ$7UZPJ65677.

\end{document}